\documentclass[twocolumn]{aastex63}

\usepackage[T1]{fontenc}

\usepackage{amsmath}
\usepackage{dcolumn}
\usepackage{hyperref}
\usepackage{gensymb}

\newcommand{\package}[1]{\textsl{#1}}

\newcommand{\paper}{\textsl{Letter}}

\newcommand{\ev}{\ensuremath{\textrm{eV}}}
\newcommand{\msun}{\ensuremath{\textrm{M}_\odot}}

\newcommand{\gyr}{\ensuremath{\textrm{Gyr}}}
\newcommand{\kpc}{\ensuremath{\textrm{kpc}}}
\newcommand{\kms}{\ensuremath{\textrm{km}\,\textrm{s}^{-1}}}

\graphicspath{{./}{figures/}}

\shorttitle{Velocity Dispersion in GD-1}
\shortauthors{Gialluca, Naidu \& Bonaca}

\begin{document}\sloppy\sloppypar\raggedbottom\frenchspacing 

\title{Velocity Dispersion of the GD-1 Stellar Stream}

\correspondingauthor{Megan~T.~Gialluca}
\email{mtg224@nau.edu}

\author[0000-0002-2587-0841]{Megan~T.~Gialluca}
\affiliation{Department of Astronomy and Planetary Science $|$ Northern Arizona University, Box 6010, Flagstaff, AZ 86011, USA}
\affiliation{Center for Astrophysics $|$ Harvard \& Smithsonian, 60 Garden Street, Cambridge, MA 02138, USA}

\author[0000-0003-3997-5705]{Rohan~P.~Naidu}
\affiliation{Center for Astrophysics $|$ Harvard \& Smithsonian, 60 Garden Street, Cambridge, MA 02138, USA}

\author[0000-0002-7846-9787]{Ana~Bonaca}
\affiliation{Center for Astrophysics $|$ Harvard \& Smithsonian, 60 Garden Street, Cambridge, MA 02138, USA}

\begin{abstract}\noindent
Tidally dissolved globular clusters form thin stellar streams that preserve a historical record of their past evolution.
We report a radial velocity dispersion of $2.3\pm0.3\,\kms$ in the GD-1 stellar stream using a sample of 43 spectroscopically confirmed members.
The GD-1 velocity dispersion is constant over the surveyed $\approx15\degree$ span of the stream.
We also measured velocity dispersion in the spur adjacent to the main GD-1 stream, and found a similar value at the tip of the spur.
Surprisingly, the region of the spur closer to the stream appears dynamically colder than the main stream.
An unperturbed model of the GD-1 stream has a velocity dispersion of $\approx0.6\,\kms$, indicating that GD-1 has undergone dynamical heating.
Stellar streams arising from globular clusters, which prior to their arrival in the Milky Way, orbited a dwarf galaxy with a cored density profile are expected to have experienced the amount of heating required to match the velocity dispersion observed in GD-1.
This suggests that GD-1 has been accreted and that imprints of its original host galaxy, including the inner slope of its dark-matter halo, remain observable in the stream today.
\end{abstract}

\keywords{stars: kinematics and dynamics -- Galaxy: kinematics and dynamics -- dark matter}

\section{Introduction} \label{sec:intro}
Due to two-body interactions, stars evaporate from globular clusters and form long, thin stellar streams \citep{Spitzer87, Combes99}.
Streams that orbit in the Galactic halo can remain coherent for billions of years \citep{Balbinot18}.
Intrinsically, they are dynamically cold, so any gravitational anomaly they encounter leaves a record in the distribution of stream stars.
Large perturbations, like the passage of a massive dark-matter subhalo, typically produce prominent underdensities, or gaps, in a stellar stream \citep{Carlberg09, Yoon11}.
Close encounters of less massive objects would predominantly scatter stream stars, thus increasing its thickness and velocity dispersion \citep{Ibata02, Johnston02}.

Excitingly, signatures of dynamical perturbation have recently been detected in the GD-1 stellar stream.
\citet{Grillmair06} discovered GD-1 as a long stellar stream without a progenitor, however its small width and small spread in metallicity \citep[][]{Bonaca20}, signal that GD-1 is a completely dissolved globular cluster.
Using proper motions from the Gaia Data Release 2 catalog \citep{gaiadr2} to confidently identify stream member stars, \citet{pwb} discovered large density variations along the GD-1 stream, as well as stars beyond the main stream.
These structures were also found using deep photometry alone \citep{deBoer18}.
The origin of these features is unclear, as similar features are produced in simulations of streams that have had a close encounter with a compact, massive object, like a low-mass dark-matter subhalo \citep[e.g.,][]{Bonaca19, Banik20}, simulations of an unperturbed stream with a much more massive progenitor \citep[e.g.,][]{Ibata20}, and in simulations of stellar streams accreted into the Milky Way from a smaller satellite galaxy \citep[e.g.,][]{Malhan20}.

\begin{figure*}
\centering
\includegraphics[width=0.8\textwidth]{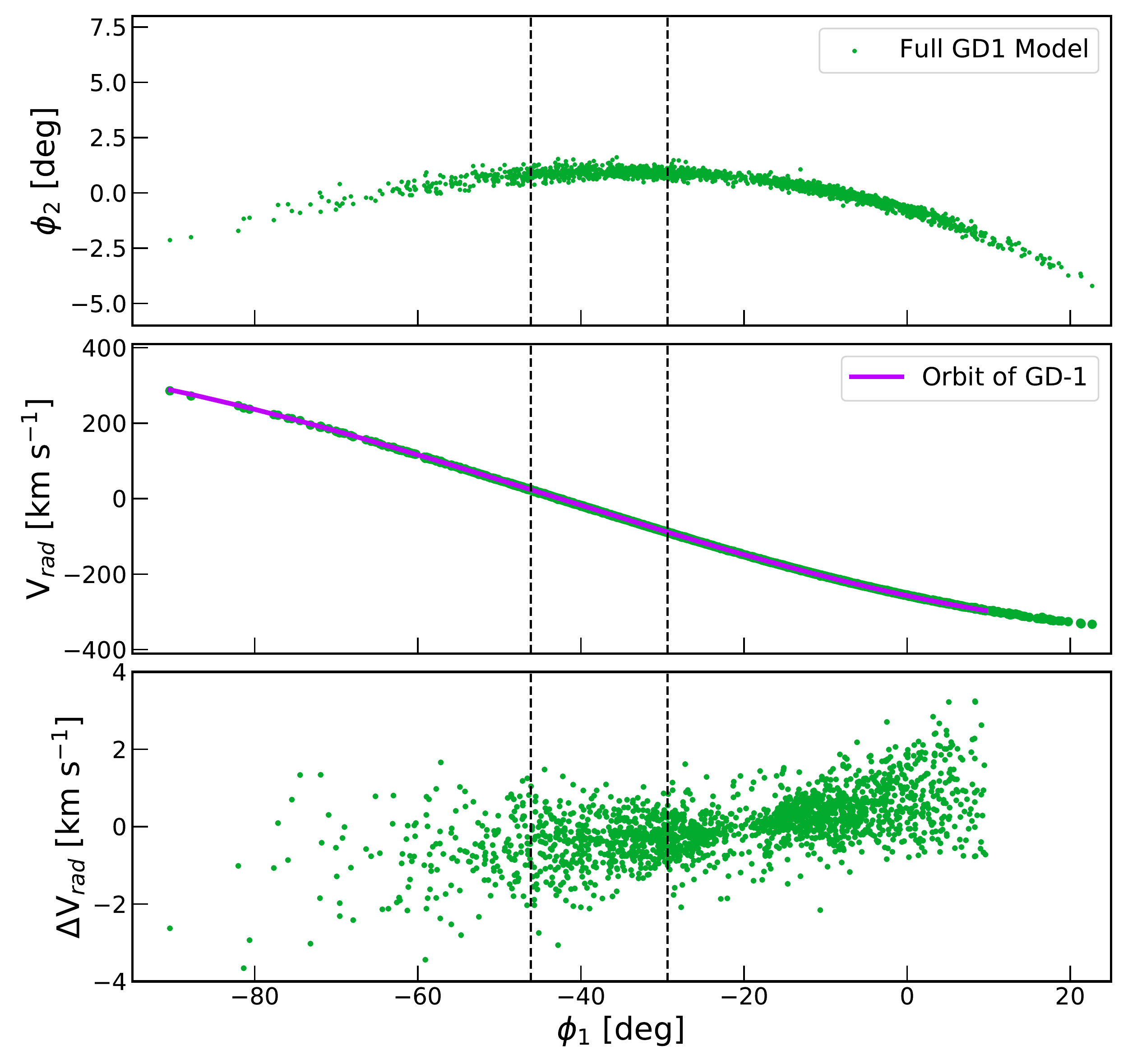}
\caption{
Top: Sky positions of our fiducial GD-1 stream model that matches the extent of the observed stream (in the stream coordinates with longitude $\phi_1$ and latitude $\phi_2$).
Middle: Radial velocity, $V_{\rm rad}$, along the stream.
The purple line shows the orbital radial velocity of the modeled GD-1, $V_{\rm rad,orb}$.
Bottom: Relative radial velocity of the modeled stream with respect to the orbital radial velocity, $\Delta V_{\rm rad} = V_{\rm rad}-V_{\rm rad,orb}$.
We analyzed the model between dashed black lines, which enclose the region where high-resolution spectroscopy is available for GD-1 stars.
}
\label{fig:CloseSim}
\end{figure*}

Velocity dispersion serves as an important indicator of the cumulative amount of perturbation a stream has experienced, and would therefore help to distinguish the relative importance of external and internal processes in shaping the GD-1 stellar stream.
Radial velocities are available for a large number of stream member stars distributed over the entire extent of GD-1 \citep{Koposov10, Huang19}, but due to their large measurement uncertainties, they only put an upper limit on the radial velocity dispersion in GD-1 of $\sigma_{vrad}\leq3\,\kms$.
Using Gaia proper motions, \citet{Malhan19} also put an upper limit on the average tangential velocity dispersion across GD-1 of $\sigma_{vtan}\leq2.3\,\kms$.
Recently, \citet{Bonaca20} published a catalog of GD-1 members distributed over a limited range along the stream, but observed with a high-resolution spectrograph so that radial velocity uncertainties are precise enough to resolve a dispersion in radial velocity as low as $\approx1\,\kms$.

In this \paper, we first explore how the velocity dispersion of a stellar stream on the GD-1 orbit depends on its age and the progenitor's mass, and estimate the velocity dispersion GD-1 would have in the absence of any perturbations (\S\,\ref{sec:sims}).
In \S\,\ref{sec:sigma} we use radial velocities from \citet{Bonaca20} to determine velocity dispersion in GD-1, and show it is significantly larger than predicted by unperturbed models.
This suggests that GD-1 has undergone dynamical heating, and in \S\,\ref{sec:discussion} we explore plausible heating sources.

\begin{figure}
\centering
\includegraphics[width=\columnwidth]{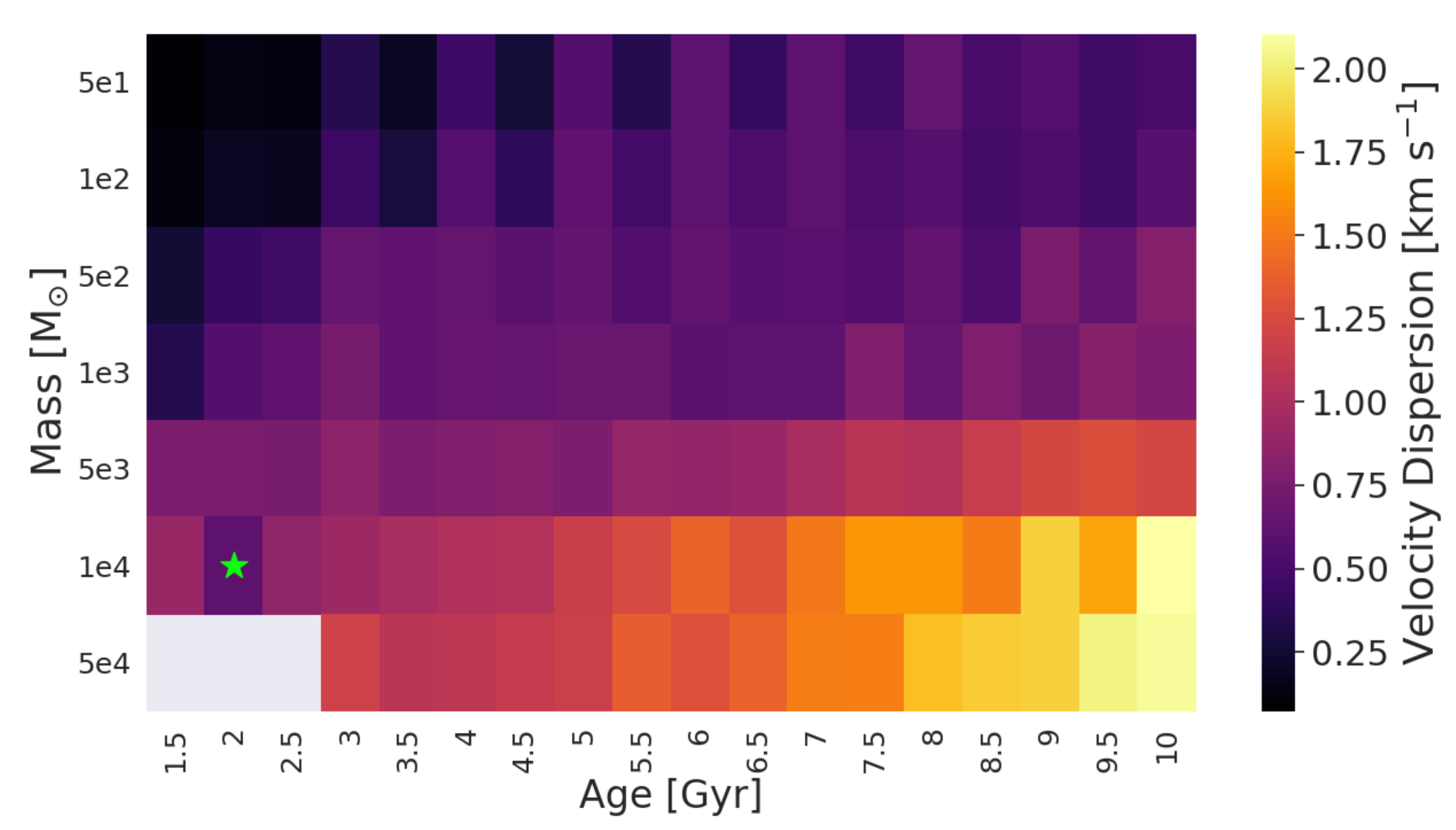}
\caption{
Velocity dispersion in simulated GD-1 streams as a function of progenitor mass and stream age.
Green star represents our fiducial GD-1 model whose velocity dispersion is $0.6\,\kms$ (see Figure \ref{fig:CloseSim}).
In general, velocity dispersion increases with both progenitor mass and stream age, with the mass dependence being stronger than age.}
\label{fig:SimSummary}
\end{figure}

\section{Intrinsic velocity dispersion in GD-1}
\label{sec:sims}
To determine the velocity dispersion GD-1 would have in the absence of dynamical perturbations, we created a suite of stream models on the GD-1 orbit covering a range of progenitor masses and stream ages.
Because the stream is narrow \citep{pwb}, has little dispersion in metallicity \citep{Bonaca20} and no visible progenitor, we assume the GD-1 progenitor is a completely disrupted globular cluster.
We simulate the stream by releasing tracer particles from the progenitor cluster's tidal radius and evolving them in a Milky Way gravitational potential \citep[the streakline method,][]{Bonaca14, Fardal15}.
Our Milky Way model includes a \citet{miyamoto1975three} disk (mass of $5.5\times10^{10}\,\msun$, scale length of 3 kpc, and scale height of 0.28 kpc), a \citet{hernquist1990analytical} bulge (mass of $4\times10^9\,\msun$, and scale radius of 1\,kpc), and a \citet{navarro1997universal} dark-matter halo (scale mass of $7\times10^{11}\,\msun$, scale radius of 15.62\,kpc, and a flattening of 0.95).

We found that a model with the progenitor's initial mass $M_{\rm init} = 10^4\,\msun$ evolved for $\tau=2\,\gyr$, in which the progenitor disrupted after $t_{\rm dis}=1\,\gyr$ matches well the observed extent of the stream (Figure~\ref{fig:CloseSim}, top panel, shown in the $\phi_{1,2}$ stream coordinate system).
The simulated stream shows a strong radial velocity gradient (Figure~\ref{fig:CloseSim}, middle panel), which is well described by the orbital radial velocity (purple line).
Radial velocity relative to the orbital, $\Delta V_{\rm rad}$, shows that the stream is kinematically cold (Figure~\ref{fig:CloseSim}, bottom panel).
In the region where precise velocities are available observationally ($-46\degree \leq\phi_1\leq -29\degree$), the fiducial unperturbed model of the GD-1 stellar stream has a velocity dispersion of $\sigma_{V_r}=0.6\,\kms$. 

To further explore the range of velocity dispersion a stream like GD-1 can attain unperturbed, we varied the progenitor's initial mass and the stream age.
For a globular cluster on a given orbit, the dissolution time depends mainly on its initial mass and is given by:
\begin{equation}
\label{eq:tdis}
t_{\rm dis} = t_4\left(\frac{M_{\rm init}}{10^4\,\msun}\right)^\gamma
\end{equation}
where $t_4$ is the disruption time of a $10^4\,\msun$ globular cluster.
We used values of the power law index $\gamma=0.62$ and $t_4=1\,\gyr$, appropriate for clusters dissolving in the inner Milky Way \citep{boutloukos2003star}.
Based on Equation \ref{eq:tdis}, the mass cannot exceed $10^7\,\msun$ as that would cause the disruption time to be greater than the age of the universe.
Because observations of GD-1 show no progenitor, the progenitor disruption time must be shorter than the stream age.
Considering these constraints, we ran simulations for progenitor masses of $50\,\msun - 5\times10^4\,\msun$ and for stream ages of $1.5\,\gyr - 10\,\gyr$.
Figure \ref{fig:SimSummary} shows the velocity dispersion as a function of the initial mass and age.
Our fiducial model is marked with a green star in this Figure.
Dispersion increases with increasing mass and age, with mass being the more influential parameter.
However, for most combinations of the GD-1's progenitor mass and stream age the expected velocity dispersion remains low, and only approaches $\approx2\,\kms$ for the two oldest ($\tau\geq9.5\,\gyr$) and most-massive models ($M_{\rm init}=5\times10^4\,\msun$) in our grid.
With no external perturbations, a stream like GD-1 should be very cold dynamically.

\section{Measured velocity dispersion in GD-1}
\label{sec:sigma}
\begin{figure*}
\centering
\includegraphics[width=\textwidth]{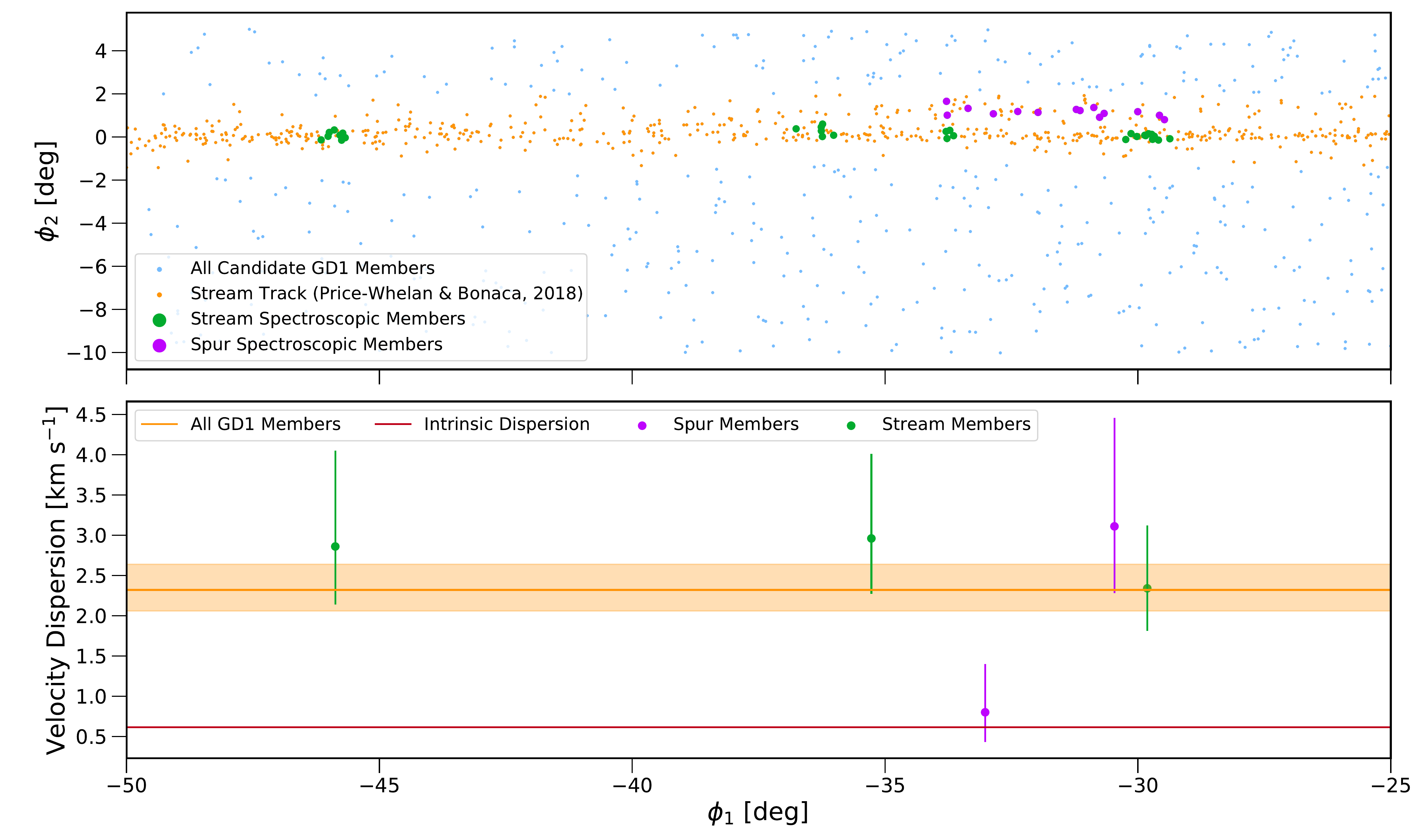}
\caption{
Top: sky positions of likely GD-1 stars (orange points), and spectroscopically confirmed members in the main stream (green) and in the spur (purple).
Bottom: Velocity dispersion along the GD-1 stream.
Velocity dispersion measured in GD-1 overall (orange shaded region) is significantly higher than expected for an unperturbed stream on this orbit (dark red line).
}
\label{fig:VelocityDispersions}
\end{figure*}

We use precise radial velocities of 43 dynamically and chemically identified GD-1 members from \citet{Bonaca20} to measure the stream's velocity dispersion.
The top panel of Figure~\ref{fig:VelocityDispersions} shows the sky locations of these stars in the GD-1 coordinate system defined by \citet{Koposov10}, with $\phi_{1,2}$ being the stream longitude and latitude, respectively.
Our sample contains stars located both in the main GD-1 stream (green points) and in the spur (purple points).
The observed radial velocities show a strong gradient along the stream, driven by the orbital velocity trend.
Here we consider radial velocities relative to the orbital radial velocity, $\Delta V_{\rm rad} = V_{\rm rad} - V_{\rm rad, orbit}$, where $V_{\rm rad}$ is the measured radial velocity, and $V_{\rm rad, orbit}$ is the orbital radial velocity derived as a function of the stream longitude in \citet{Bonaca20}.

\begin{figure}
\centering
\includegraphics[width=\columnwidth]{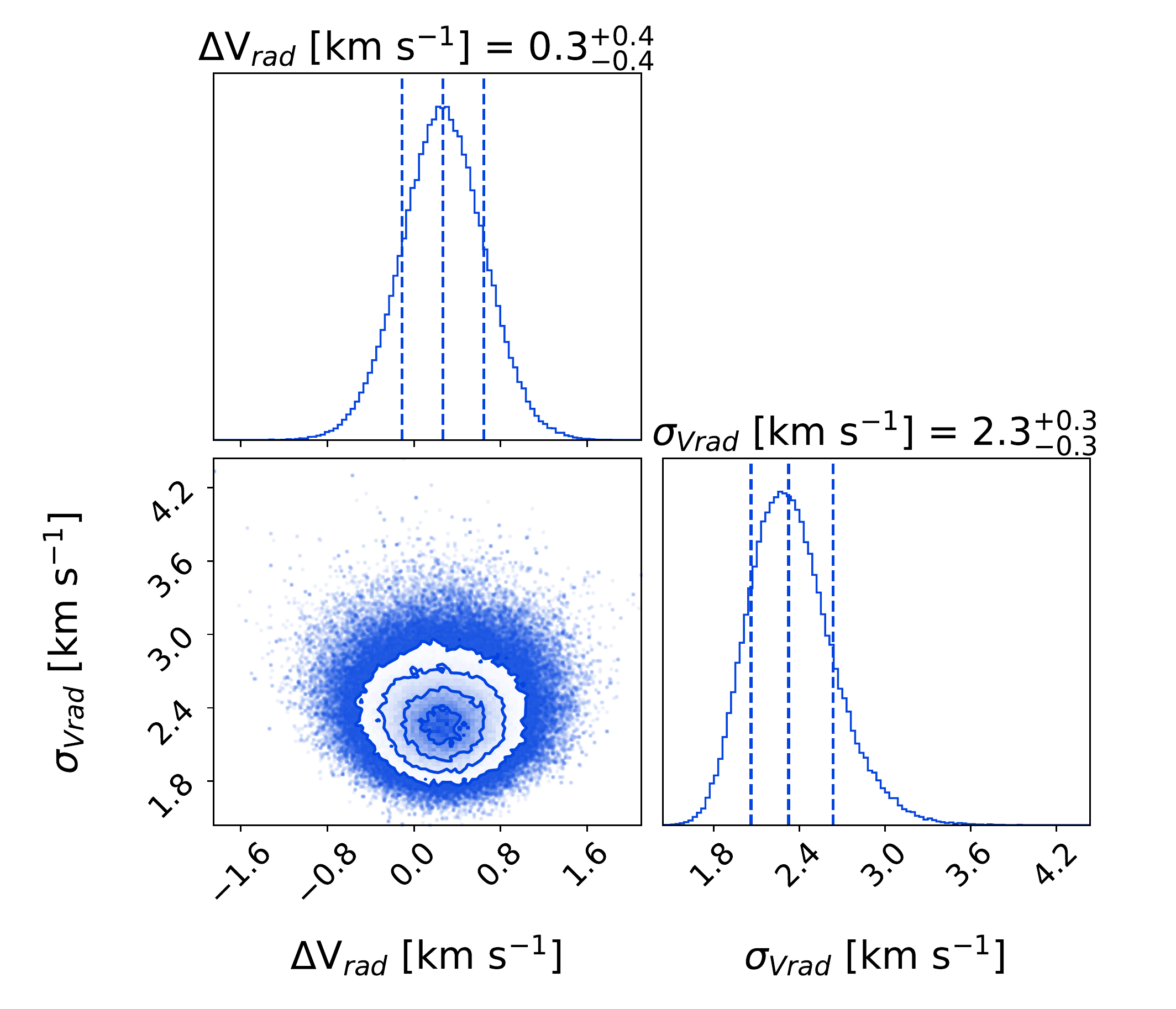}
\caption{
Posterior probability distributions of the relative radial velocity ($\Delta V_{\rm{rad}}$) with respect to the stream's orbit and the stream's velocity dispersion ($\sigma_{\rm{Vrad}}$) constrained by all GD-1 member stars.
}
\label{fig:Distributions}
\end{figure}

We assume that the relative radial velocities, $\Delta V_{\rm rad,i}$ are normally distributed around the best-fit orbit (parameterized by $\mu$), and that both the observational uncertainties, $\sigma_i$, and the stream velocity dispersion, $\sigma_{\rm Vrad}$, contribute to the overall dispersion, $\Sigma$, such that $\Sigma^2 = \sigma_i^2 + \sigma_{\rm Vrad}^2$.
Therefore, the probability of relative radial velocity of a star $i$ given the mean radial velocity offset and dispersion is:
\begin{equation}
p_i(\Delta V_{\rm rad,i}|\mu, \Sigma) = \mathcal{N}(\Delta V_{\rm rad,i}|\mu, \Sigma)
\end{equation}
Assuming that the radial velocity measurements of individual stars are independent, the joint likelihood is simply $p (\{\Delta V_{\rm rad}\}|\mu, \Sigma) = \prod_i \mathcal{N}(\Delta V_{\rm rad,i}|\mu, \Sigma)$.
We measured the posterior distribution of the mean relative radial velocity and dispersion in GD-1 by sampling this likelihood using the affine invariant Markov Chain Monte Carlo ensemble sampler \package{emcee} \citep{emcee}.
We advanced 200 walkers for 2500 steps, and analyzed the converged chains after discarding the first 500 burn-in steps.
Posterior distributions of the relative radial velocity and velocity dispersion for the GD-1 stream overall are shown in Figure~\ref{fig:Distributions}.
As expected, the relative offset of radial velocity measurements from the best-fit orbit is small and consistent with zero.
The stream dispersion, however, is well resolved at $\sigma_\textrm{Vrad}=2.3\pm0.3\,\kms$ which is considerably larger than expected for an unperturbed GD-1 stream (see Section~\ref{sec:sims}).

The bottom panel of Figure~\ref{fig:VelocityDispersions} shows velocity dispersion along the GD-1 stream.
Our sample contains stars observed in eight fields, some of which have only a handful of member stars.
To mitigate the effect of small sample size on the estimate of the stream's velocity dispersion, adjacent low-density fields have been combined so that every measurement is based on at least 6 stars.
With the exception of an extremely cold spur field at $\phi_1\approx-33\degree$, which has a low dispersion of $\sigma\lesssim1\,\kms$, velocity dispersion in the observed stream region is remarkably constant at $\sigma\approx3\,\kms$.

\section{Summary \& Discussion}
\label{sec:discussion}
We measured the average radial velocity dispersion of $\sigma_{\rm Vrad}=2.3\pm0.3\,\kms$ in a region of the GD-1 stellar stream that contains the adjacent spur ($-46\degree\lesssim\phi_1\lesssim-29\degree$).
Velocity dispersion is approximately constant along this part of the stream, except in the inner part of the spur which is somewhat colder than the average.
We produced a fiducial numerical model of the GD-1 stream and found that unperturbed, it has an intrinsic velocity dispersion of $\sigma_\textrm{intrinsic}=0.6\,\kms$.
On a given orbit, dispersion is larger for older streams and for streams with more massive progenitors.
However, to match the observed velocity dispersion, we would need an unrealistically massive progenitor and dynamically old stream.
Therefore, the observed velocity dispersion implies GD-1 has been dynamically heated.

The morphology of the gap and the spur observed in the GD-1 stream can be reproduced by numerical models of the stream interacting with a massive object in the Milky Way halo \citep{Bonaca19, deBoer20}, and the interaction might have elevated the stream's velocity dispersion.
However, similarly high values of velocity dispersion were measured in the Palomar~5 tidal tails \citep[$\sigma=2.1\pm0.4\,\kms$,][]{Kuzma15} and the ATLAS-Aliqa Uma complex \citep[$\sigma=4.8\pm0.4\,\kms$,][]{Li20}.
This motivates us to explore stream heating mechanisms that can affect many streams globally.

\subsection{Fuzzy dark matter heating}
When simulating stellar streams, we represented the Milky Way's dark matter halo as a spatially smooth, analytic density distribution.
However, most models of dark matter predict that dark matter halos have at least some amount of substructure \citep[][]{Springel08, Bode01}.
In the case of cold dark matter, low-mass subhalos orbiting within the Milky Way would heat streams like GD-1 \citep[e.g.,][]{Ibata02, Johnston02}.
As these subhalos have yet to be detected, dark matter models that severely suppress substructure on small scales have been proposed.
Fuzzy dark matter is one such model where dark matter is an ultra-light axion of mass $\approx10^{-22}\,\ev$ \citep{Hu00}.
Due to the low particle mass, fuzzy dark matter exhibits quantum effects on astrophysical scales, including fluctuations in the local density of dark matter.
Numerical simulations show that this quantum turbulence can dynamically heat stellar streams \citep{Amorisco18}, as well as field stars \citep{Church19}.
Here we test whether fuzzy dark matter heating can explain the velocity dispersion observed in GD-1.

Following \citet{Amorisco18}, we assume that the fuzzy dark matter fluctuations can be modeled as a population of soliton clumps whose density is given by the local dark matter density, and whose size, $r_{\rm sc}$, increases for decreasing axion mass: $r_{\rm sc}=r_{\rm sc,1}/m_{22}$, where $m_{\rm FDM}=m_{22}\times10^{-22}\,\ev$ is the axion mass and $r_{\rm sc,1}=0.2\,\kpc$ the soliton radius appropriate for the Milky Way halo.
In our model of the Milky Way, average dark matter density along the GD-1 orbit is $\rho=4.2\times10^6\,\msun\,\kpc^{-3}$.
So, the effective mass of soliton clumps that GD-1 encounters is:
\begin{equation}
M_{\rm eff} = 4\pi/3\,r_{\rm sc}^3\,\rho = 1.4\times10^5 m_{22}^{-3}\,\msun
\end{equation}
As expected, for lighter axion particles the effective mass of soliton clumps increases and so does dynamical heating due to fuzzy dark matter turbulence.

Encounter with a soliton clump imparts a velocity kick to stream stars.
Averaged over time, these velocity kicks increase the stream's velocity dispersion.
\citet{Hui17} derived an expression for the velocity dispersion increase as a function of the perturbers' mass, size, and number density (Equation~48), which we integrate over time to obtain cumulative velocity dispersion due to soliton clump heating:
\begin{equation}
\sigma_{\rm FDM}^2 = \frac{4 \pi (k G M_{\rm eff} w)^2 N t}{V r_{\rm sc}^2}
\label{eq:sigma_fdm}
\end{equation}
where $N=1/V_{\rm sc}=3/(4\pi)r_{\rm sc}^{-3}$ is the number density of clumps, $V$ is the soliton speed relative to the stream, $w$ is the stream width and $t$ its age, $G$ is the gravitational constant, and $k$ is a geometric factor.
For simplicity, we assume $k=1$ (appropriate if soliton clumps were point masses).
Substituting definitions for the effective mass and size of soliton clumps in Equation~\ref{eq:sigma_fdm}, we obtain the following expression for the axion mass as a function of stream velocity dispersion:
\begin{equation}
m_{\rm FDM} = \frac{3t\, (Gw\,1.4\times10^5\,\msun)^{2}}{V \sigma_{FDM}^2 r_{sc,1}^5}\times10^{-22}\,\ev
\label{eq:m22}
\end{equation}
As found by \citet{Amorisco18}, dynamically hotter streams imply lighter fuzzy dark matter particles.

Assuming that the velocity dispersion we observed in GD-1 is entirely due to fuzzy dark matter heating, i.e., $\sigma_{\rm FDM}=2.3\,\kms$, that the solitons are moving at typical halo speeds $V=220\,\kms$, adopting the stream age from our fiducial model $t=2\,\gyr$, and taking the stream width from literature \citep[$w=FWHM=0.5\degree$, or 80\,pc at 10\,\kpc][]{Koposov10}, we find $m_{\rm FDM}=4.1^{+1.1}_{-0.9}\times10^{-24}\,\ev$.
To take into account that unperturbed stellar streams have non-zero velocity dispersion, we next assume that the GD-1 velocity dispersion due to fuzzy dark matter heating is $\sigma_{\rm FDM} = \sqrt{\sigma_{\rm GD-1}^2 - \sigma_{\rm fid}^2}=2.1\,\kms$, where $\sigma_{\rm Vrad}$ is the measured velocity dispersion (\S\,\ref{sec:sigma}), and $\sigma_{\rm fid}$ is the velocity dispersion in the fiducial GD-1 model (\S\,\ref{sec:sims}), yielding only a slightly more massive estimate $m_{\rm FDM}=4.4^{+1.3}_{-1.0}\times10^{-24}\,\ev$.
Fuzzy dark matter particles this light have been ruled out by the population of low-mass dwarf galaxies in the Milky Way \citep[$m_{\rm FDM}>2.9\times10^{-21}\,\ev$,][]{Nadler20}.
Heating from the allowed fuzzy dark matter models is insufficient to produce velocity dispersion measured in this part of the GD-1 stellar stream.

\subsection{Progenitor galaxy heating}
One important piece of context is that GD-1 likely arrived with a dwarf galaxy.
In recent years the stellar halo, and a number of halo globular clusters, have been shown to arise almost entirely out of accretion \citep{Massari19,Naidu20}.
This accretion origin likely also holds for disrupted globular clusters like GD-1.
Locating the host dwarf galaxy of GD-1 and dating its accretion redshift, through e.g., ages of its main-sequence stars \citep{Bonaca20b}, or dynamical arguments \citep{Koppelman19}, would further clarify the origin of the stream. 

Understanding this origin is important, since GD-1 may have undergone some degree of processing in its accreted host galaxy prior to falling into the Milky Way, adding to its present day observed velocity dispersion \citep{Carlberg18, Carlberg20}.
Following the disruption of globular clusters that were accreted into the Milky Way with a dwarf galaxy, \citet{Malhan20} found that the velocity dispersion of resulting streams depends on the dark matter density profile of their original host galaxies.
For cuspy dwarf galaxy hosts, the stream velocity dispersion is expected in the range $\sigma_{vrad}\approx5-10\,\kms$, significantly higher than that of GD-1.
However, cored dwarfs are less disruptive to their globular clusters, and as a result their streams are colder with $\sigma_{vrad}\approx2-3\,\kms$.
The radial velocity dispersion we measured in GD-1 can be explained by the GD-1 progenitor cluster having been accreted in a cored dwarf galaxy.

While tantalizing, the accretion origin of the GD-1 stream is still a preliminary finding.
In this work we only measured radial velocities in a fraction of the stream, which may have an elevated dispersion due to a local perturbation (e.g., one that produce the stream's gap and spur).
Similarly precise measurements across the entire stream are required to establish whether the velocity dispersion is globally high, as expected if GD-1 were accreted.
Independently, tidal debris of the host galaxy should be identifiable as a broader structure on similar orbits in the Galactic halo, whose velocity dispersion is also sensitive to the inner slope of dark matter density \citep{Errani15}.

Our study of velocity dispersion in the GD-1 stellar stream is but a teaser of what will soon be possible in many streams.
Massive spectroscopic surveys like SDSS-V \citep{sdss5} and DESI \citep{desi} are slated to deliver millions of stellar spectra.
Similar analyses of these data will allow us to measure structural properties in a population of dissolved Milky Way progenitors and test dark matter models through velocity dispersions of stellar streams orbiting in the Galactic halo.

\vspace{0.5cm}
The SAO REU program is funded in part by the National Science Foundation REU and Department of Defense ASSURE programs under NSF Grant no.\ AST-1852268, and by the Smithsonian Institution.
We would also like to acknowledge Dr. Matthew Ashby and Dr. Jonathan McDowell for their guidance, assistance, and support.
We further thank Ali Kurmus, Diana Khimey, Victoria Ono, and Sownak Bose for valuable feedback and support during the summer group meetings.
AB acknowledges support from NASA through HST grant HST-GO-15930.

\software{
\package{Astropy} \citep{astropy, astropy:2018},
\package{emcee} \citep{emcee},
\package{gala} \citep{gala},
\package{matplotlib} \citep{mpl},
\package{numpy} \citep{numpy},
\package{scipy} \citep{scipy}
}

\bibliography{Ref}{}
\bibliographystyle{aasjournal}

\end{document}